# COVID-19 Modeling Based on Real Geographic and Population Data

Emir BAYSAZAN [1], A. Nihat BERKER [2,3,4], Hasan MANDAL [5], Hakan KAYGUSUZ [6,7,*]

[1] TEBIP High Performers Program, Board of Higher Education of Turkey, Istanbul University, Fatih, Istanbul 34452, Turkey

[2] Faculty of Engineering and Natural Sciences, Kadir Has University, Cibali, Istanbul 34083, Turkey

[3] TÜBİTAK Research Institute for Fundamental Sciences, Gebze, Kocaeli 41470, Turkey

[4] Department of Physics, Massachusetts Institute of Technology, Cambridge, Massachusetts 02139, USA

[5] The Scientific and Technological Research Council of Turkey (TÜBİTAK), Kavaklıdere, Ankara 06100, Turkey

[6] Department of Basic Sciences, Faculty of Engineering and Architecture, Altınbaş University, Istanbul 34217, Turkey.

[7] SUNUM Nanotechnology Research Center, Sabancı University, Tuzla, Istanbul 34956, Turkey

**\*Correspondence:** hakan.kaygusuz@altinbas.edu.tr

**Abstract:** Intercity travel is one of the most important parameters for combating a pandemic. The ongoing COVID-19 pandemic has resulted in different computational studies involving intercity connections. In this study, the effects of intercity connections during an epidemic such as COVID-19 are evaluated using a new network model. This model considers the actual geographic neighborhood and population density data. This new model is applied to actual Turkish data by the means of provincial connections and populations. A Monte Carlo algorithm with a hybrid lattice model is applied to a lattice with 8,802 data points. Results show that this model is quantitatively very efficient in modeling real world COVID-19 epidemic data based on populations and geographical intercity connections, by the means of estimating the number of deaths, disease spread, and epidemic termination.

## 1. Introduction

The Coronavirus disease 2019 (COVID-19) pandemic is caused by the severe acute respiratory syndrome coronavirus 2 (SARS-CoV-2). The use of several vaccination measures such as Sinovac CoronaVac, Sputnik V, Pfizer-BioNTech BNT162b2 and Turkovac has reduced the effects of the pandemic. However, it is still considered to be ongoing and has caused societal impacts, including in urban areas. The complexity of the pandemic has been compounded by the interconnected nature of societal activities.

During the pandemic, indication to vaccine development (Mandal 2021), many countries adapted different methods to combat this disease. These include improved hygiene measures, vaccination, curfews and full quarantines. Turkey is among the countries which applied full quarantines followed by curfews on weekends. Among the precautions, intercity travel was restricted and subject to special permission as one of the measures to combat disease transmission given the spatial diffusion patterns through intercity mobility (Gu et al. 2022).

Many scientists and research groups are studying with in silico methods of genetic analysis (Ugurel et al. 2020, Eskier et al. 2021). Computation methods also include the epidemiological simulation. Two main approaches followed in the literature are fitting the clinical data with various mathematical models and the simulation of the epidemic with various mathematical models (Maltezos and Georgakopoulou 2021). Previously, several studies on modeling infectious diseases have been reported (Filipe and Gibson 1998, 2001, Draief and Ganesh 2011, Bestehorn et al. 2021, Ahmetolan et al. 2022). Among the simulation methods, Monte Carlo (MC) simulation is one of the effective methods (Xie 2020). Currently, various MC studies are present for COVID-19 simulation. These include analyzing different scenarios for selected countries (Vyklyuk et al. 2021), age-structured mobility data for simulation of the pandemic spread in selected cities (De Sousa et al. 2020), random-walk proximity-based infection spread (Triambak and Mahapatra 2021, Mahapatra and Triambak 2022) and observing the effect of weekend curfews (Kaygusuz and Berker 2021).

Travel restrictions are among the first emergency measures (Tian et al. 2020) during epidemics. Inter-city travel models on epidemic spreading can be modeled using the well-known small-world networks (Watts and Strogatz 1998) and recent papers report such simulations (An et al. 2021). Big data analytics have been used to compare mobility patterns during the pandemic in Finland (Kiashemshaki et al. 2022) while other studies have focused on the impacts of the pandemic on public transport (Naveen and Gurtoo 2022). Moreover, the most recent report of the Intergovernmental Panel on Climate Change underlined the impacts of the pandemic on society, including the transport sector (IPCC 2022).

This paper discusses the effects of intercity travel during a pandemic such as COVID-19 using a new model network on a selected country (Turkey) by considering the actual geographic neighborhood and population density data. The spread of the disease was modeled using a hybrid of susceptible, infected, quarantine and recovered (SIQR) (Kermack and Mckendrick 1927, Huppert and Katriel 2013, Shu et al. 2016) lattice model (Liccardo and Fierro 2013) and spin-1 Ising model (Berker and Wortis 1976, Hoston and Berker 1991), which is previously reported by our recent paper (Kaygusuz and Berker 2021). The present study addresses the gap in the literature for coupling actual geographic neighborhood and population density data with the a hybrid lattice model.

## 2. Methods

The Monte Carlo model used in this study is written in the Python language by the authors. A lattice model with 8,802 data points was generated using the map of Turkey, taking into consideration the neighborhoods of provincial centers and population densities. Figure 1A shows the lattice model based on the provincial centers of Turkey and Figure 1B indicates the population-weighted version of the one-to-one same lattice, where provinces with more population are represented with bigger lattice blocks. The lattice shown in Figure 1B is considered in further simulations.

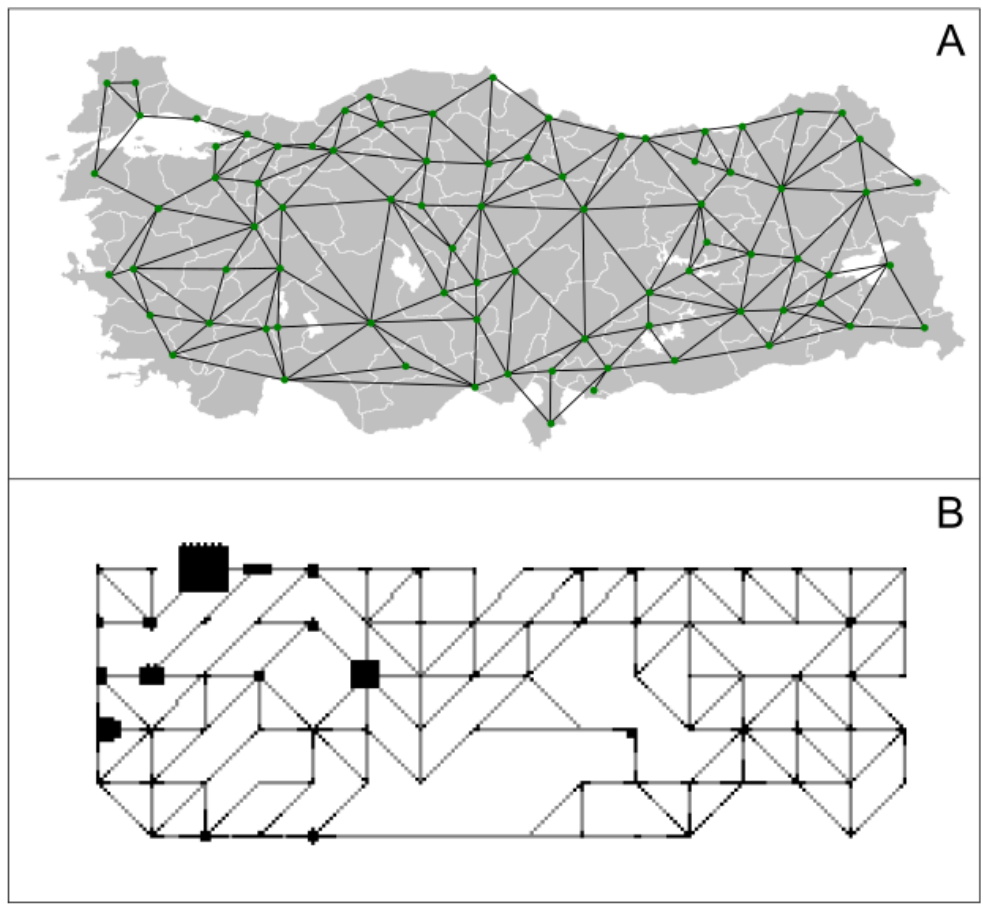

**Figure 1.** Lattice model based on the provincial neighborhoods overlaid on the map of Turkey (A) and population weighted lattice of the same lattice (B).

Provinces with bigger population are shown in bigger lattice blocks, where the actual population data was obtained from the Turkish Statistical Institute (Turkish Statistical Institute 2021). The network is generated on an Ising-like model where the persons are placed on lattice points. The number of people in each block (city) in the matrix has been assigned according to their ratio in the total Turkish population. On the border of each town there are roads that provide intercity connections. These are designed as connections between square-shaped population blocks. The connections are unpopulated; however, they can transmit the disease to a person in the neighboring town. For each person in the network, five different states are possible: Healthy (susceptible), positive, sick (under quarantine), recovered, and deceased. Unpopulated parts in the total lattice are assigned as forbidden lattice points.

For the Monte Carlo simulation, persons in the lattices are randomly selected and assigned a random possible direction for the movement. Firstly, neighbors (i.e., side neighbors including intercity connections and corner neighbors on the square lattice) are checked for contamination with probability $P$. Secondly, the selected person moves in the selected direction with probability $P$. If the selected direction contains a person, this happens by person interchange.

After all persons in the lattice are considered for one turn, this is defined as one Monte Carlo step. The simulation considers the following scenario: At step 0, a random person gets positive (patient zero) in Istanbul (since it is the largest city with arguably the largest cosmopolitan connections) and starts spreading the disease. Once a person gets into contact with the virus, the person can remain in quarantine with a probability ($P_{quarantine} = 0.33$) or be recovered with a probability ($P_{recovery} = 0.33$) at each Monte Carlo Step. A sick person under quarantine cannot move and cannot spread the disease, and recovers with probability ($P_{treatment} = 0.995$) or is deceased with probability $1-P_{treatment}$, after 7 Monte Carlo steps. This ratio is arranged according to total mortality rate of COVID-19 in Turkey (~0.005). For each person, a random number is generated in the script and if this random number is bigger than the probability $P$, the related event occurs. In this model, no lockdowns are considered.

## 3. Results and Discussion

Figure 2 shows the virus spread in Bursa on days 0 and 50, respectively. Persons with the disease are marked with red.

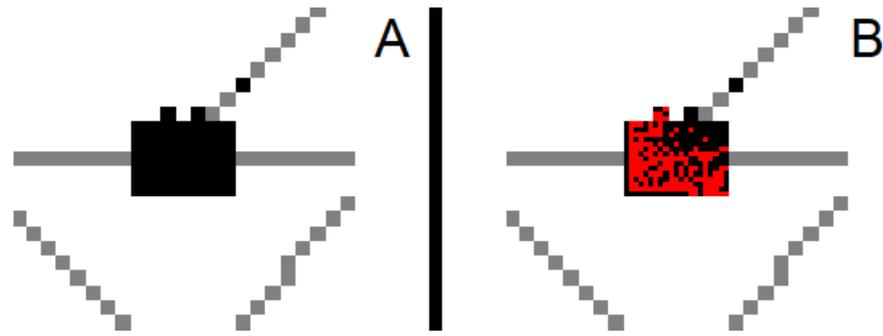

**Figure 2.** Spread of the disease in Bursa a) day 0, b) day 50

The result of the epidemic is tested using the susceptible, infected, in quarantine and recovered (SIQR) model, which is previously applied for COVID-19 (Odagaki 2020, Odagaki 2021).

Figure 3 shows the active positive cases in the simulation. Around MC step 70, the number of active cases in Turkey reaches up to 8.0% of the total population, followed by a second wave at around MC step 100, around 7.0%. After this second wave, the number of active cases decreases and reaches down to 0 around MC step 160.

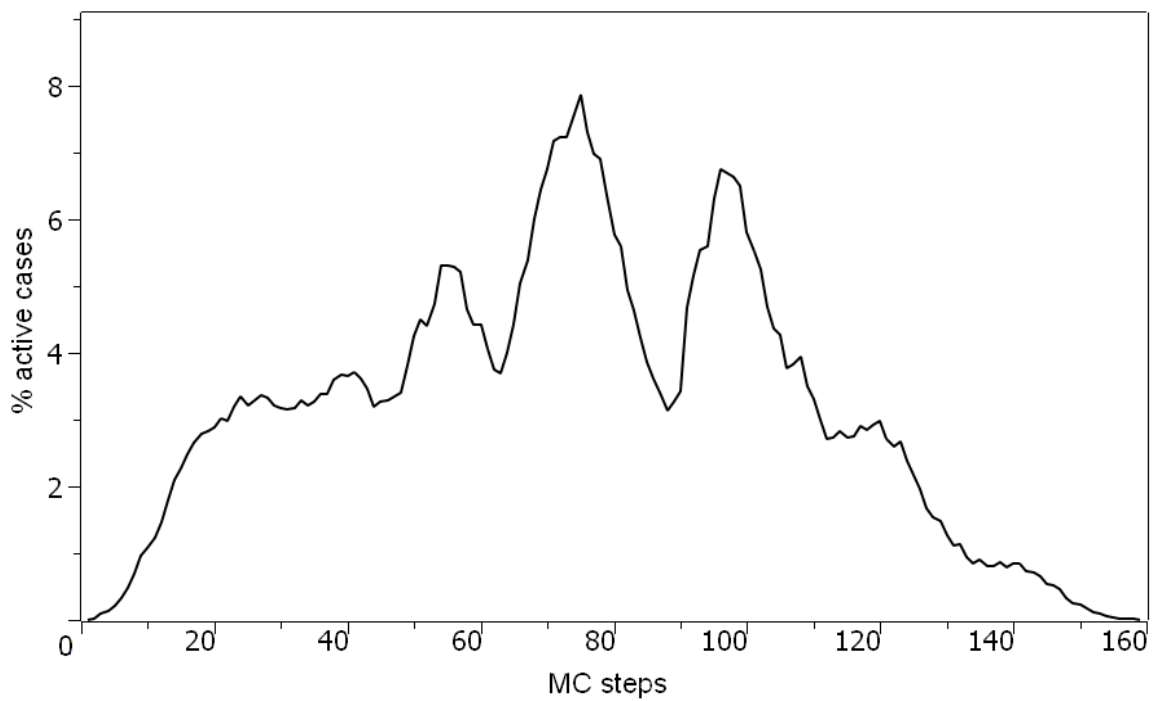

**Figure 3.** Number of active cases (as a percentage of population) in this model.

Among the different epidemic measures, the primary goal is to reduce the number of deaths. The number of deaths for Turkey from the beginning of the pandemic is compared with the simulation data and is shown in Figure 4, by the means of day-percentage of death to the total population. Real-world data was obtained from Worldometer (Worldometer 2022). Results show that the simulation fits real-world data well and can be used as an efficient tool to predict the number of deaths. Another important outcome is that ten MC steps of the simulation are around one real-time day. This implies that the population density-based lattice of Turkish provinces is effective in modeling the real data. Numbers of healthy and susceptible individuals are also provided in Figure 5.

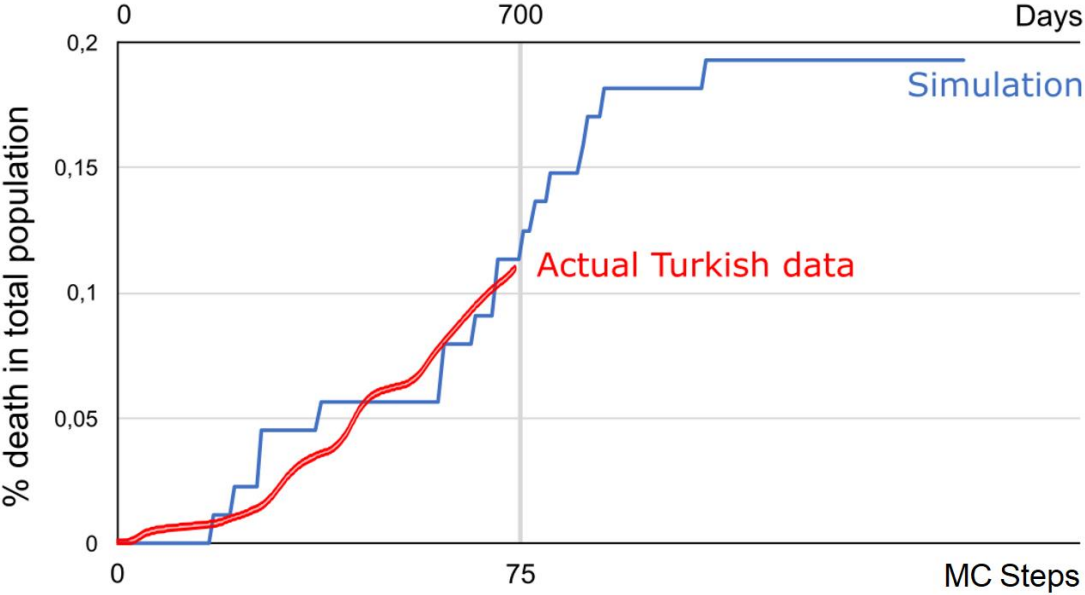

**Figure 4.** Percentage of deaths due to COVID-19, simulation compared with real Turkish data.

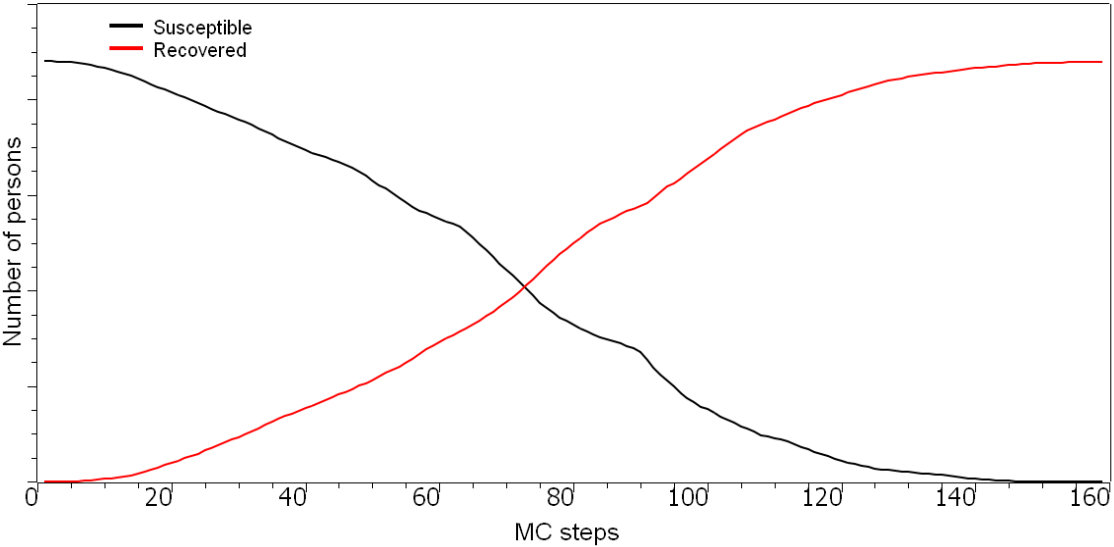

**Figure 5.** Number of healthy (susceptible) and recovered individuals over MC steps

In this model, the infectivity and the mortality of the virus are constant, which means there are no variants of the initial virus. In addition, vaccination is neglected, which assumes that the entire population has the same level of immunity to the virus. Under these circumstances, the epidemic in Turkey extinguishes on 160 MC steps. Currently the pandemic continues but daily active cases are about to run out around the date article is written.

Another subject examined in this study is the spread of diseases in different provinces of Turkey. In order to visualize the spread of the disease, the percentage of positive cases (quarantined and non-quarantined) in the province was evaluated for each block (province) and the results for 81 Turkish provinces are shown in Figure 6. Starting with Istanbul, the epidemic quickly expands into most provinces between MC steps 60-100. Seven selected provinces from seven geographical regions of Turkey are also examined for highlighting the spread of epidemic in different regions (in Figure 7). It is shown that the spread vanishes at the Southeastern Anatolia Region around MC step 160.

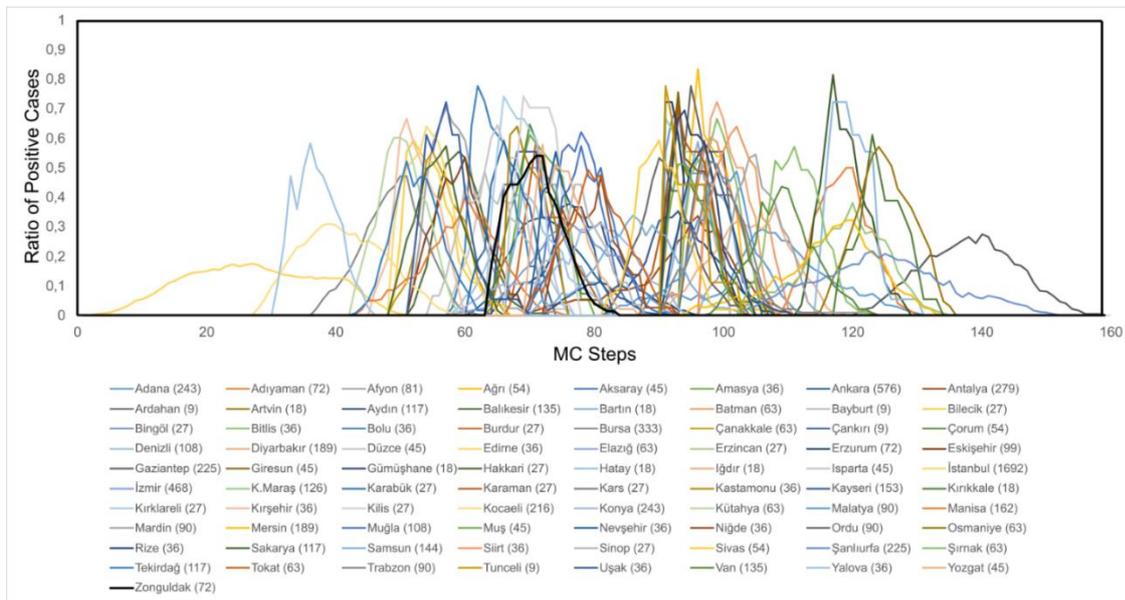

**Figure 6.** Spread of the epidemic in 81 Turkish provinces (names and number of persons in each province are shown below the plot).

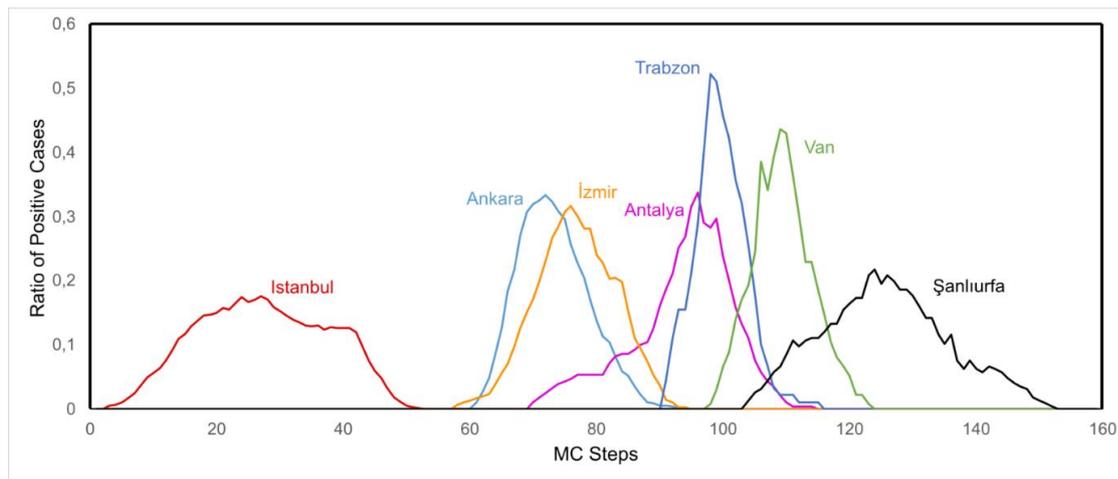

**Figure 7.** Spread of the epidemic in seven cities of seven geographical regions of Turkey.

In conclusion, in this study, a novel approach to COVID-19 epidemic simulation was conducted and tested on epidemic data in Turkey. This model uses the provincial neighborhood and population data. A hybrid Monte Carlo algorithm ensured a very efficient modeling of the epidemic by the means of growth time and number of deaths. Further studies can reveal this model in different countries, as well as the effects of different prevention measures, such as different types of vaccines, travel bans and curfews. The coupling of spatiotemporal datasets with the hybrid lattice model is promising of numerous other advances in the field.

**Acknowledgments**

We thank F. Bedia Erim for a critical reading of our manuscript. A. Nihat Berker gratefully acknowledges support by the Academy of Sciences of Turkey (TÜBA).

**References**

Ahmetolan S, Bilge AH, Demirci A, Dobie AP (2022). A Susceptible–Infectious (SI) model with two infective stages and an endemic equilibrium. Mathematics and Computers in Simulation 194: 19–35. doi: 10.1016/J.MATCOM.2021.11.003.

An Y, Lin X, Li M, He F (2021). Dynamic governance decisions on multi-modal inter-city travel during a large-scale epidemic spreading. Transport Policy 104: 29–42. doi: 10.1016/J.TRANPOL.2021.01.008.

Berker AN, Wortis M (1976). Blume-Emery-Griffiths-Potts model in two dimensions: Phase diagram and critical properties from a position-space renormalization group. Physical Review B 14: 4946–4963. doi: 10.1103/PhysRevB.14.4946.

Bestehorn M, Riascos AP, Michelitsch TM, Collet BA (2021). A Markovian random walk model of epidemic spreading. Continuum Mechanics and Thermodynamics: 1–15. doi: 10.1007/s00161-021-00970-z.

De Sousa LE, Neto PHDO, Da Silva Filho DA (2020). Kinetic Monte Carlo model for the COVID-19 epidemic: Impact of mobility restriction on a COVID-19 outbreak. Physical Review E 102: 032133. doi: 10.1103/PhysRevE.102.032133

Draief M, Ganesh A (2011). A random walk model for infection on graphs: Spread of epidemics & rumours with mobile agents. Discrete Event Dynamic Systems 21: 41–61. doi: 10.1007/s10626-010-0092-5.

Eskier D, Akalp E, Dalan Ö, Karakülah G, Oktay Y (2021). Current mutatome of sars-cov-2 in turkey reveals mutations of interest. Turkish Journal of Biology 45: 104–113. doi: 10.3906/biy-2008-56.

Filipe JAN, Gibson GJ (1998). Studying and approximating spatio-temporal models for epidemic spread and control. Philosophical Transactions of the Royal Society B: Biological Sciences 353: 2153–2162. doi: 10.1098/rstb.1998.0354.

Filipe JAN, Gibson GJ (2001). Comparing approximations to spatio-temporal models for epidemics with local spread. Bulletin of Mathematical Biology 63: 603–624. doi: 10.1006/bulm.2001.0234.

Gu L, Yang L, Wang L, Guo Y, Wei B, Li H (2022). Understanding the spatial diffusion dynamics of the COVID-19 pandemic in the city system in China. Social Science and Medicine 302: 114988. doi: 10.1016/j.socscimed.2022.114988.

Hoston W, Berker AN (1991). Multicritical phase diagrams of the blume-emery-griffiths model with repulsive biquadratic coupling. Physical Review Letters 67: 1027–1030. doi: 10.1103/PhysRevLett.67.1027.

Huppert A, Katriel G (2013). Mathematical modelling and prediction in infectious disease epidemiology. Clinical Microbiology and Infection 19: 999–1005. doi: 10.1111/1469-0691.12308.

IPCC (2022). Climate Change 2022: Mitigation of Climate Change [online]. Website: https://www.ipcc.ch/report/ar6/wg3/ [accessed 01 July 2022].

Kaygusuz H, Berker AN (2021). The effect of weekend curfews on epidemics: a Monte Carlo simulation. Turkish Journal of Biology 45: 436–441. doi: 10.3906/biy-2105-69.

Kermack WO, Mckendrick AG (1927). A contribution to the mathematical theory of epidemics. Proceedings of the Royal Society of London. Series A, Containing Papers of a Mathematical and Physical Character 115: 700–721. doi: 10.1098/rspa.1927.0118.

Kiashemshaki M, Huang Z, Saramäki J (2022). Mobility Signatures: A Tool for Characterizing Cities Using Intercity Mobility Flows. Frontiers in Big Data 5: 822889. doi: 10.3389/fdata.2022.822889.

Liccardo A, Fierro A (2013). A Lattice Model for Influenza Spreading. PLoS ONE 8: e63935. doi: 10.1371/journal.pone.0063935.

Maltezos S, Georgakopoulou A (2021). Novel approach for Monte Carlo simulation of the new COVID-19 spread dynamics. Infection, Genetics and Evolution 92: 104896. doi: 10.1016/j.meegid.2021.104896.

Mandal H (2021). Achievements of the COVID-19 Turkey Platform in vaccine and drug development with an approach of "co-creation and succeeding together." Turkish Journal of Medical Sciences 51: 3139–3149. doi: 10.3906/sag-2112-178.

Naveen BR, Gurtoo A (2022). Public transport strategy and epidemic prevention framework in the Context of Covid-19. Transport Policy 116: 165-174. doi: 10.1016/j.tranpol.2021.12.005.

Shu P, Wang W, Tang M, Zhao P, Zhang YC (2016). Recovery rate affects the effective epidemic threshold with synchronous updating. Chaos 26: 063108. doi: 10.1063/1.4953661.

Tian H, Liu Y, Li Y, Wu CH, Chen B, et al. (2020). An investigation of transmission control measures during the first 50 days of the COVID-19 epidemic in China. Science 368: 638–642. doi: 10.1126/SCIENCE.ABB6105.

Triambak S, Mahapatra DP (2021). A random walk Monte Carlo simulation study of COVID-19-like infection spread. Physica A: Statistical Mechanics and its Applications 574: 126014. doi: 0.1016/j.physa.2021.126014.

Turkish Statistical Institute (2021). Address Based Population Registration System Results [online]. Website: https://data.tuik.gov.tr/Bulten/Index?p=Adrese-Dayali-Nufus-Kayit-Sistemi-Sonuclari-2021-45500 [accessed: 14 May 2022].

Ugurel OM, Ata O, Turgut-Balik D (2020). An updated analysis of variations in SARS-CoV-2 genome. Turkish Journal of Biology 44: 157–167. doi: 10.3906/biy-2005-111.

Vyklyuk Y, Manylich M, Škoda M, Radovanović MM, Petrović MD (2021). Modeling and analysis of different scenarios for the spread of COVID-19 by using the modified multi-agent systems – Evidence from the selected countries. Results in Physics 20: 103662. doi: 10.1016/j.rinp.2020.103662.

Watts DJ, Strogatz SH (1998). Collective dynamics of 'small-world' networks. Nature 393:6684 393: 440–442. doi: 0.1038/30918.

Worldometer (2022). Turkey COVID - Coronavirus Statistics [online]. Website: https://www.worldometers.info/coronavirus/country/turkey/ [accessed 14 May 2022].

Xie G (2020). A novel Monte Carlo simulation procedure for modelling COVID-19 spread over time. Scientific Reports 10: 1–9. doi: 10.1038/s41598-020-70091-1.